\documentclass{article}
\usepackage{spconf,amsmath,graphicx}

\usepackage{graphicx}
\usepackage{multirow}
\usepackage{booktabs}
\usepackage{amsmath}
\usepackage{amssymb}
\usepackage{adjustbox}

\usepackage{enumitem}
\setlist{nosep, leftmargin=14pt}

\usepackage{mwe} 

\title{Deep Semi-supervised Metric Learning with Dual Alignment for Cervical Cancer Cell Detection}
\name{Anonymous ISBI2022 submission}
\address{Paper ID 236}

\name{Zhizhong Chai$^{1}$, Luyang Luo$^{2}$, Huangjing Lin$^{1,2}$, Hao Chen$^{3}$, Anjia Han$^{4}$, Pheng-Ann Heng$^{2}$}

\address{$^{1}$ Imsight AI Research Lab
\\$^{2}$ Dept. Computer Science \& Engineering, The Chinese University of Hong Kong
\\$^{3}$ Dept. Computer Science \& Engineering, The Hong Kong University of Science and Technology \\
$^{4}$ Department of Pathology, The First Affiliated Hospital, Sun Yat-sen University}

\begin{document}
%
\maketitle
\begin{abstract}
Deep learning has achieved unprecedented success in various object detection tasks with huge amounts of labeled data.
However, obtaining large-scale annotations for medical images is extremely challenging due to the high demand of labour and expertise.
In this paper, we propose a novel deep semi-supervised metric learning method to effectively leverage both labeled and unlabeled data for cervical cancer cell detection.
Specifically, our model learns a metric space and conducts dual alignment of semantic features on both the proposal level and the prototype levels.
On the proposal level, we align the unlabeled data with class proxies derived from the labeled data.
We further align the prototypes of the labeled and unlabeled data to alleviate the influence of possibly noisy pseudo labels generated at the proposal alignment stage.
Moreover, we adopt a memory bank to store the labeled prototypes, which significantly enrich the metric learning information from larger batches.
Extensive experiments show our proposed method outperforms other state-of-the-art semi-supervised approaches consistently, demonstrating the efficacy of our proposed deep semi-supervised metric learning with dual alignment.
\end{abstract}
\section{Introduction}
\label{sec:intro}

Cervical cancer screening is an effective way to prevent the occurrence rate of cervical cancer. 
Digitalization and artificial intelligence (AI) technology could empower computational pathology with the potential to alleviate cytologists from the overburdened workload \cite{lin2021dual,wang2019weakly,xie2019deep}. Nevertheless, to the best of our knowledge, most of the cervical lesion detection methods are fully supervised, lacking the consideration of massive unlabeled data.
How to fulfill the potential and benefit from unlabeled data with massive hard mimics is challenging yet of great value for cervical cancer detection.

Recently, many semi-supervised object detection methods have been proposed under different application scenarios.
A broad branch of works was based on knowledge distillation.
For example, Wang et al. \cite{wang2020focalmix} and Luo et al. \cite{Luo2021OXnet} utilized the soft target focal loss to take care of the foreground-background imbalance when learning from ensembled predictions. 
Zhou et al. \cite{zhou2020deep} proposed consistency constraints on both the logits and feature maps with a teacher model.
These methods highly rely on the quality of predictions generated by self-ensembling models or prediction ensembles.
Another branch of works generates pseudo labels for the unlabeled data.
Liu et al. \cite{Liu2021Unbiased} proposed to utilize focal loss to better learn the pseudo labels.
Wang et al. \cite{wang2021co} presented the co-mining algorithm that combines the predictions from the Siamese network for refining pseudo labels.
Nevertheless, these works could be prone to overfitting on the pseudo labels.

To enable large-scale semi-supervised cervical cancer cell detection, we present a novel deep semi-supervised metric learning network.
Our model performs metric learning with both proposal-level and prototype-level alignment in a metric space to learn more discriminative features.
On the proposal level, we generate pseudo labels for the unlabeled data to align the proposal features with learnable categorical proxies derived from the labeled images.
As the pseudo labels are possibly noisy, we further propose to align the labeled and unlabeled prototypes generated from each mini-batch of the training data.
Moreover, we adopt a memory bank to store the labeled prototypes and hence enrich the metric learning information from larger batches.
To the best of our knowledge, metric learning studies are often on fully labeled data \cite{deng2019arcface} or purely unlabeled data \cite{he2020momentum}. Here, we explore to further unleash the potential of metric learning models by unifying both labeled and unlabeled data.
Extensive experiments show our proposed method improves the fully-supervised baseline under various scenarios and outperforms other state-of-the-art semi-supervised detection approaches.

\label{sec:method}

\begin{figure*}[!t]
\center
\includegraphics[width=0.85\textwidth]{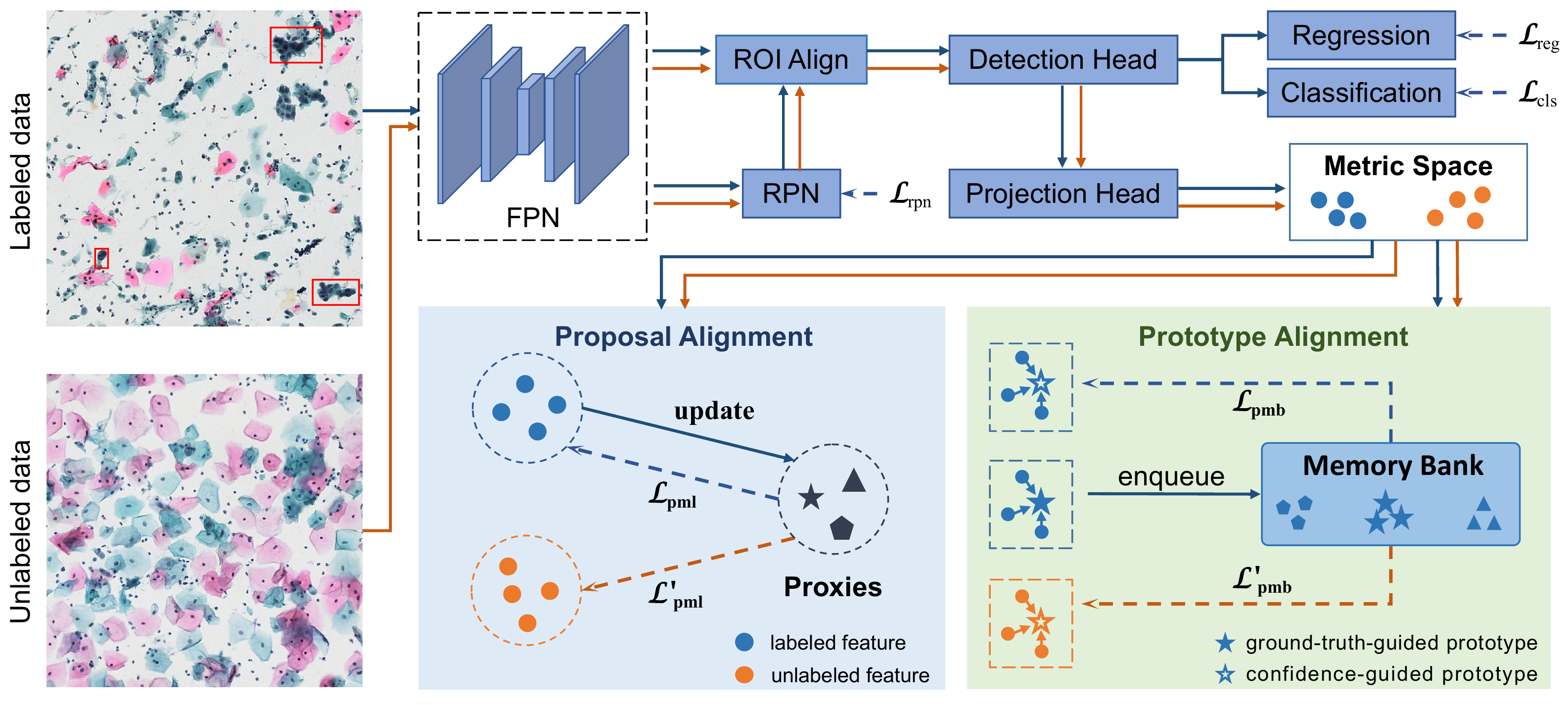}
\caption{Overview of our proposed framework. For supervised learning, the standard supervised losses are calculated to optimize the model. For unsupervised learning, we conduct the proposal alignment and prototype alignment in the same metric space generated by the projection head.}
\label{fig_framework}
\end{figure*}

\section{Methodology}

Let $\mathcal{D}_{L}$ denote the labeled dataset and $\mathcal{D}_{U}$ denote the unlabeled dataset. 
The goal of semi-supervised cervical cancer detection is improving the performance by effectively leveraging $\mathcal{D}_{U}$ without human annotating labor. 
Here, We construct our semi-supervised metric learning framework based on the Faster R-CNN \cite{ren2016faster} with Feature Pyramid Network (FPN) \cite{lin2017feature}.
Fig. \ref{fig_framework} illustrates the overview of our proposed method, which will be elaborated in the following sections.

\subsection{Proxy-based Proposal Alignment}
\label{sec:proposal alignment}

Instead of adding extra training objectives onto the classifiers as in existing literature \cite{wang2020focalmix, zhou2020deep, Liu2021Unbiased, wang2021co}, we conduct metric learning on a learned metric space for semantic alignment of the proposal features of the labeled and unlabeled data.
Formally, let $H$ denote the proposal feature generated by the detection head, we first incorporate a projection head to map the proposal features into the metric space:
\begin{equation}
x = W^{2}\sigma(W^{1}H)
\end{equation}
where $W^{1}$ and $W^{2}$ denote two fully connected layers for dimension reduction, and $\sigma$ is the ReLU function \cite{nair2010rectified}.
Next, we adopt the learnable semantic proxies as the class representatives, which are defined as a part of the network parameters that can be optimized through training \cite{Movshovitz2017No}. To alleviate the negative influence of pseudo labels from unlabeled data, we only update the proxies from the labeled data. Denoting $p_{c}$ as the proxy belongs to the $c$-th category, for labeled data, the proxy-based metric learning loss is defined as:
\begin{equation}
    \mathcal{L}_{pml} = -\log{\frac{\exp(-\Phi(x_{c},p_{c}))}{\sum_{k\neq c}^{C} \exp(-\Phi(x_{c},p_{k}))}}
\end{equation}

\noindent where $x_{c}$ denotes an embedding belonging to category $c$ out of in total $C$ classes, and $\Phi(\cdot,\cdot)$ represents a similarity measurement function computing the Euclidean distance. For unlabeled data, we use pseudo labels generated by the model for each proposal embedding. The proxy-based loss for unlabeled data is as follows:
\begin{equation}
  \mathcal{L}_{pml}' = -\log{\frac{\exp(-\Phi(\tilde{x_{c}},p_{c}))}{\sum_{k\neq c}^{C} \exp(-\Phi(\tilde{x_{c}},p_{k}))}}.
\end{equation}
where $\tilde{x_{c}}$ is the embeddings for unlabeled data, of which the label is determined by thresholding the network's prediction. We could now successfully align both the labeled and unlabeled proposal embeddings with the proxies. 

\subsection{Memory Bank-based Prototype Alignment}
\label{sec:prototype alignment}
Pseudo labeling is a hard assumption that could easily result in overfitting on model's own predictions.
To enable smoother regularization, we adopt the prototype alignment loss to further maintain the semantic consistency between the labeled and unlabeled data.
Specifically, the mean features of labeled proposals for each class are computed as, namely, ground truth-guided prototypes $\mathcal{P}^{gg}$.
We then aggregate proposals with a confidence-based fusion strategy to be, namely, confidence-guided prototypes $\mathcal{P}^{cg}$, which can alleviate the influence of low-confidence samples as suggested in \cite{xu2020cross}.
Formally, the prototypes are obtained as follows:
\begin{equation}
   \mathcal{P}^{gg} =\frac{1}{N}\sum_{i=1}^{N}{x}_{i} \text{   , and      } \mathcal{P}^{cg} = \frac{\sum_{i=1}^{N}{q}_{i} \cdot {x}_{i}}{\sum_{i=1}^{N}{q}_{i}}
\end{equation}
\noindent where $N$ is the number of proposals (could be labeled or unlabeled) in a mini batch, and $q_i$ is the prediction score for the proposal projection $x_i$. Note that we separately compute $\mathcal{P}_{L}^{cg}$ and $\mathcal{P}_{U}^{cg}$ for labeled data and unlabeled data.
We then align the two types of prototypes in the metric space:

\begin{equation}
\mathcal{L}_{intra}(P^{cg},P^{gg})=\frac{\sum_{i=1}^{C}\Phi(P_{c}^{cg},P_{c}^{gg})}{C}
 \end{equation}
 
 
\begin{equation}
\begin{aligned}
   &\mathcal{L}_{inter}(P^{cg},P^{gg})=\frac{\sum_{c=1}^{C}\sum_{k \neq c}^{C}max(0, m -\Phi(P_{c}^{cg},P_{k}^{gg}))}{C\cdot(C-1)}
  \end{aligned}
\end{equation} 
where m is the margin term which is set to $1.0$ in our experiments. 
However, the generated prototypes of labeled and unlabeled images in one mini-batch may have the problem of category mismatch. In other words, some classes of the labeled data in one mini-batch may not exist in the unlabeled data, so we can not conduct the prototype alignment for these classes. Therefore, we further adopt a memory bank \cite{wu2018unsupervised} to store the ground-truth-guided prototypes ($P_{L}^{gg}$) from previous steps, hence effectively enriching the metric learning information from larger batches.
Moreover, in order to enhance the semantic consistency of the prototypes from training data, we conduct the alignment loss between the confidence-guided prototypes (including labeled data $P_L^{cg}$ and unlabeled data $P_U^{cg}$) and the mean prototypes of the memory bank. Denote the mean prototypes of the memory bank as $P^{M}$, we construct prototype-level alignment as follows:
\begin{small}

 \begin{equation}
 \mathcal{L}_{pmb} = \mathcal{L}_{intra}(P^{M},P_{L}^{cg})+\mathcal{L}_{inter}(P^{M},P_{L}^{cg})+\mathcal{L}_{inter}(P_{L}^{cg},P_{L}^{cg})
 \end{equation}

\begin{equation}
\begin{aligned}
\mathcal{L}_{pmb}' = \mathcal{L}_{intra}({P}^{M},P_{U}^{cg})+\mathcal{L}_{inter}(P^{M},P_{U}^{cg})+\mathcal{L}_{inter}(P_{U}^{cg},P_{U}^{cg})
\end{aligned}
\end{equation}
\noindent where $\mathcal{L}_{pmb}$ denotes losses computed by the labeled data, and $\mathcal{L}_{pmb}'$ denotes losses computed by the unlabeled data.
\subsection{Optimization and Implementation Details}
The final training loss is defined as follows:
  \begin{equation}
 \mathcal{L}_{total} = \mathcal{L}_{s}+\lambda(t)(\mathcal{L}_{pml}+\mathcal{L}_{pml}'+\mathcal{L}_{pmb}+\mathcal{L}_{pmb}')
 \end{equation}
 \end{small}
\noindent where $\mathcal{L}_{s}=\mathcal{L}_{rpn}+\mathcal{L}_{cls}+\mathcal{L}_{reg}$, which is the conventional supervised loss of Faster R-CNN. $\lambda(t)$ is a loss weighting function following typical semi-supervised learning methods \cite{jeong2019consistency,2017Mean} for stable training.
We use Faster-RCNN\cite{ren2016faster} with ImageNet-pre-trained ResNet-101 \cite{he2016deep} backbone and FPN \cite{lin2017feature} as our base model. 
All experiments are conducted based on Pytorch \cite{paszke2017automatic}. During training, we use 2 TITAN Xp GPUs with a batch size of 8, where the labeled data and unlabeled data are equally sampled. For proposal alignment, the confidence threshold used for the pseudo label is 0.5, the proxies are randomly initialized and updated by Stochastic Gradient Descent (SGD) with an initial learning rate of 0.01. For prototype alignment, the size of the memory bank is set to 1024. Random flip and color jittering are applied for data augmentation.
The model weights are updated by SGD with a momentum of 0.9, and the initial learning rate is 0.005 and multiplied by 0.1 every 40000 iterations.
\section{Experiments and Results}

\subsection{Dataset and Evaluation Metrics}
A large-scale cervical pathology dataset was collected with 240,860 images in total, containing lesions of undetermined significance (ASC-US), low-grade squamous intraepithelial lesion (LSIL), high-grade squamous intraepithelial lesion (HSIL), and atypical glandular cells (AGC). Among all data, 42,073 images with sizes 1200$\times$1200 were conducted by the liquid-based Pap test specimens from 997 patients and used as the labeled data.
These images were divided into the labeled training set, validation set, and testing set with a ratio of 7:1:2 without overlapping of patients. Then, the fully supervised detection network (Faster R-CNN) was trained to screen regions of interest from the whole slide images of other 1427 patients, which generated our unlabeled dataset $D_{U}$(a total of 198,787 patches with a size of 1200x1200). Since the cervical cells are usually small, we adopted the mAP from AP10 to AP70 as the evaluation metric to better evaluate the model’s performance.

\subsection{Evaluation with Different Dataset Settings}
In this experiment, we compare the performance of our proposed method with the fully-supervised Faster R-CNN, a widely adopted fully-supervised object detection model.
We vary the number of labeled data for training our method while keeping the unlabeled data as the same amount under different settings.
Meanwhile, Faster R-CNN is provided with the same labeled data only.
The results are reported in Table~\ref{tab:tb1}.
As a reference, Faster R-CNN \cite{ren2016faster} achieves mAP of 15.7\%, 18.6\%, 24.4\%, and 25.4\% when provided with  25\%, 50\%, 75\%, and 100\% of the total labeled data, respectively. Our proposed method consistently improves the baseline under all settings, with mAP increasing of 1.3\%, 0.9\%, 1.0\%, and 1.6\%, respectively.
Moreover, our method achieves a comparable average mAP to that of Faster R-CNN with 25\% less labeled data (row 6 vs. 7), which demonstrates that our method can effectively mine the knowledge from unlabeled data and relieve the annotating labor.

\begin{table}[h]
\renewcommand\arraystretch{1.2}
    \centering
        \caption{Evaluation of the proposed method with different numbers of labeled data.}
    \setlength{\belowcaptionskip}{-10pt}
    \begin{adjustbox}{width=1. \linewidth}
    \begin{tabular}{p{1.5cm}|p{2.8cm}|p{1.3cm}p{1.3cm}p{1.3cm}p{1.3cm}p{1.3cm}}
    \toprule
      \hfil\multirow{2}{*}{ labeled } & \hfil\multirow{2}{*}{Method} & \multicolumn{5}{c}{mAP [$\%$]} \\
    \cline{3-7}
        &  &\hfil ASC-US & \hfil LSIL & \hfil HSIL &\hfil AGC & \hfil Avg. \\
        \hline

\hfil\multirow{2}{*}{ $25\%$ } &\hfil  Faster R-CNN\cite{ren2016faster}  &\hfil7.2  &\hfil23.0  & \hfil\textbf{11.8}& \hfil20.7 &\hfil15.7\\
&\hfil  ours&\hfil \textbf{7.7} &\hfil\textbf{26.4} & \hfil 11.3& \hfil\textbf{22.6} &\hfil\textbf{17.0}\\
\hline

\hfil\multirow{2}{*}{ $50\%$ } &\hfil  Faster R-CNN\cite{ren2016faster}  &\hfil\textbf{10.1}  &\hfil31.3  & \hfil16.8& \hfil16.1 &\hfil18.6\\
&\hfil  ours&\hfil 9.9 &\hfil\textbf{31.9} & \hfil \textbf{18.0}& \hfil\textbf{18.2} &\hfil\textbf{19.5}\\
\hline

\hfil\multirow{2}{*}{ $75\%$ } &\hfil  Faster R-CNN\cite{ren2016faster}  &\hfil\textbf{12.2}  &\hfil38.1  & \hfil\textbf{24.8}& \hfil22.5 &\hfil24.4\\
&\hfil  ours&\hfil 11.1 &\hfil\textbf{40.0} & \hfil 23.0& \hfil\textbf{27.3} &\hfil\textbf{25.4}\\
\hline

\hfil\multirow{2}{*}{ $100\%$ } &\hfil  Faster R-CNN\cite{ren2016faster}  &\hfil12.8  &\hfil40.9  & \hfil 24.0 & \hfil24.0 &\hfil25.4\\
&\hfil ours &\hfil\textbf{13.1}  &\hfil\textbf{42.0}  & \hfil\textbf{25.1}& \hfil\textbf{27.7} &\hfil \textbf{27.0}\\

\hline
         \toprule
    \end{tabular}
    \end{adjustbox}
    \label{tab:tb1}
\end{table}

\subsection{Comparison with Other Semi-supervised Methods}
We compare our proposed method with the widely-used Faster R-CNN \cite{ren2016faster} as well as two state-of-the-art semi-supervised object detection methods: 1) the consistency-based semi-supervised detection (CSD) \cite{jeong2019consistency} model which constraints consistent prediction on both classification and regression outputs for an image and its flipped version; and 2) the Mean Teacher \cite{2017Mean} model which constructs knowledge distillation on the classification outputs from a self-ensembled model. For a fair comparison, all the semi-supervised methods are implemented with the same backbone and trained with all labeled and unlabeled data.

\begin{figure}[!t]
\center
\includegraphics[width=1.\columnwidth]{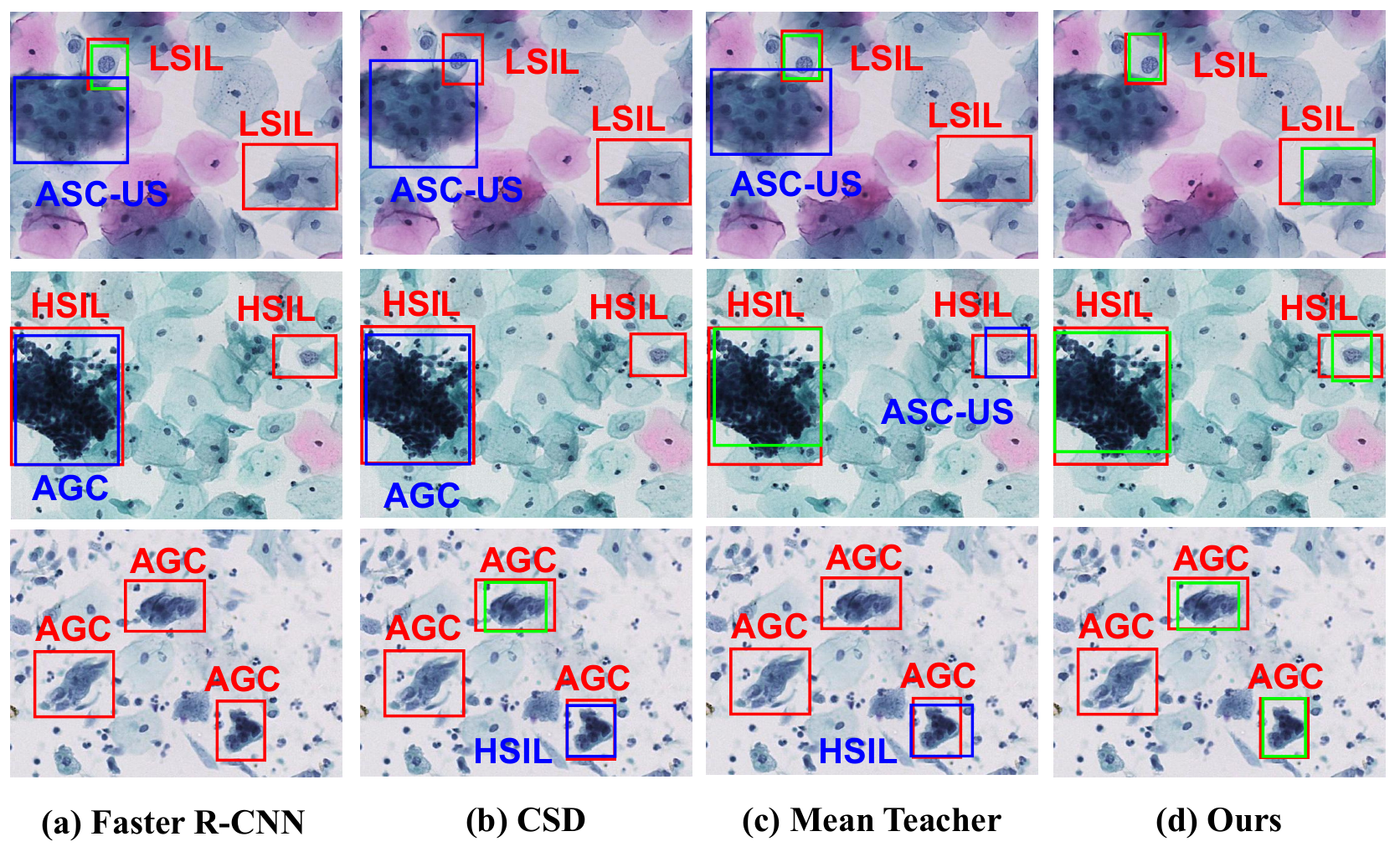}
\caption{Qualitative comparisons of semi-supervised cervical cancer cell detection on the testing set. Red rectangles stand for ground truths, green rectangles stand for true positives, and blue rectangles stand for false positives.}
\label{fig_visual_result}
\end{figure}

\textbf{Quantitative results} are reported in Table~\ref{tab:stateoftheart}. All the semi-supervised methods improve the fully supervised baseline, demonstrating effectiveness in utilizing the unlabeled data.
Compared with Faster R-CNN, the consistency constraints-based method CSD \cite{jeong2019consistency} shows a significant improvement ($6.9\%$) on AGC, while the performance on the other categories (ASC-US, LSIL, HSIL) has a certain decrease.
The knowledge distillation-based method Mean Teacher \cite{2017Mean} achieves an improvement of $1.8\%$ and $3.3\%$ on LSIL and AGC, while the performance on ASC-US and HSIL also decreases compared to the baseline model.
Notably, our proposed method shows improved performance on all the four classes from the baseline model, achieving mAP of 13.1\% in ASC-US, 42.0\% in LSIL, 25.1\% in HSIL, and 27.7\% in AGC.
Our method also achieves the best average mAP over all classes, demonstrating more efficient unsupervised data learning capability.

\textbf{Qualitative results} are illustrated in Fig.~\ref{fig_visual_result}, showing the bounding box predictions from all the compared methods. As can be observed, compared with other approaches, our proposed method yields more accurate predictions for cervical cancer cell in the liquid-based Pap images.

\begin{table}[h]
\renewcommand\arraystretch{1.2}
    \centering
        \caption{Quantitative comparisons on the test set.}
    \begin{adjustbox}{width=1. \linewidth}
    \begin{tabular}{p{5.cm}|p{1.3cm}p{1.3cm}p{1.3cm}p{1.3cm}p{1.3cm}}
    \toprule
      \hfil\multirow{2}{*}{ Method}& \multicolumn{5}{c}{mAP [$\%$]} \\
    \cline{2-6}
        &  \hfil ASC-US &\hfil LSIL & \hfil HSIL & \hfil AGC &\hfil Avg.\\
        \hline
\hfil  Faster R-CNN\cite{ren2016faster} &\hfil12.8  &\hfil40.9  & \hfil24.0& \hfil24.0 &\hfil25.4\\
\hfil  CSD \cite{jeong2019consistency} &\hfil11.6  &\hfil38.3  & \hfil22.7& \hfil\textbf{30.9} &\hfil25.9\\
\hfil  Mean Teacher \cite{2017Mean} &\hfil 11.5  &\hfil\textbf{42.7}  & \hfil23.4 & \hfil27.3 &\hfil26.2\\
\toprule
\hfil  Ours (w/o prototype alignment)&\hfil11.4  &\hfil40.0  & \hfil23.6& \hfil28.5 &\hfil25.9\\
\hfil  Ours (w/o proposal alignment)&\hfil\textbf{13.4}  &\hfil41.1  & \hfil24.0& \hfil27.7 &\hfil26.6\\
\hfil Ours &\hfil13.1  &\hfil42.0  & \hfil\textbf{25.1}& \hfil27.7 &\hfil \textbf{27.0}\\

         \toprule
    \end{tabular}
    \end{adjustbox}
    \label{tab:stateoftheart}
\end{table}

\subsection{Ablation Study of the Proposed Method}
We also conduct ablation studies to investigate the efficacy of different proposed components.
As shown in Table~\ref{tab:stateoftheart}, removing either level of alignment results in a performance decrease.
On the other hand, each alignment loss also leads to mAP improvement from the baseline Faster R-CNN.
By combining both components, our complete model achieves the best average performance.
These results demonstrate that the proposed proposal-level alignment and prototype-level alignment have complementary contributions to semi-supervised cervical cancer cell detection.

\section{Conclusion}

In this paper, we propose a novel semi-supervised deep metric learning framework with dual alignment for cervical cancer cell detection.
The proposed method learns a metric space to conduct complementary alignment on both the proposal level and prototype level.
Extensive experiments demonstrate the effectiveness and robustness of our method on the task of semi-supervised cervical cancer cell detection.
Moreover, the proposed method is general and can be easily extended to other tasks of semi-supervised object detection.

\section{Compliance with Ethical Standards}
This study was performed in line with the principles of the Declaration of Helsinki. Approval was granted by our institutional ethical committee.

\section{Acknowledgement.} 
This work was supported by the Hong Kong Innovation and Technology fund under Project ITS/170/20 and Project GHP/110/19SZ.
Zhizhong Chai and Huangjing Lin currently are employees of a company, outside this work. The other authors have no relevant financial or non-financial interest to disclose.

\bibliographystyle{IEEEbib}
\bibliography{strings,refs}

\end{document}